\documentclass[aps,prb,twocolumn,superscriptaddress,showpacs]{revtex4}
\usepackage{epsfig}
\usepackage{graphicx}
\usepackage{amsfonts}
\usepackage[figuresright]{rotating}
\usepackage{amssymb}
\usepackage{amsmath}
\usepackage{psfrag}
\def\be{\begin{equation}}
\def\ee{\end{equation}}
\def\bea{\begin{eqnarray}}
\def\eea{\end{eqnarray}}

\def\nn{\nonumber}

\begin{document}
\title{Isotropic Landau levels of Dirac fermions in high dimensions}
\author{Yi Li}
\affiliation{Department of Physics, University of California, San Diego,
CA 92093}
\author{Kenneth Intriligator}
\affiliation{Department of Physics, University of California, San Diego,
CA 92093}
\author{ Yue Yu}
\affiliation{
Institute of Theoretical Physics, Chinese Academy of Sciences, P.O. Box 2735,
Beijing 100190, China}
\author{Congjun Wu}
\affiliation{Department of Physics, University of California, San Diego,
CA 92093}
\affiliation{Center for Quantum Information, IIIS, Tsinghua University, Beijing, China}
%\date{\today}

\begin{abstract}
We generalize the Landau levels of two-dimensional Dirac fermions to three
dimensions and above with the full rotational symmetry.
Similarly to the two-dimensional case, there exists a branch of zero
energy Landau levels of fractional fermion modes
for the massless Dirac fermions.
The spectra of
other Landau levels distribute symmetrically with respect to the zero energy
scaling with the square root of the Landau level indices.
This mechanism is a non-minimal coupling of Dirac fermions
to the background fields.
This high dimensional relativistic Landau level problem is a square root
problem of its previous studied non-relativistic version investigated in Li and Wu [arXiv:1103.5422 (2011)].
\end{abstract}
\pacs{73.43.-f,71.70.Ej,75.70.Tj}
\maketitle

\section{Introduction}
\label{sect:intro}
The integer quantum Hall effect in two-dimensional (2D) electron gas arises
from the quantized 2D Landau levels (LL).
The non-trivial band structure topology is characterized by non-zero
Chern numbers \cite{thouless1982,kohmoto1985}.
Later on, quantum anomalous Hall insulators based on Bloch-wave
band structures were proposed in the absence of Landau
levels \cite{haldane1988}.
In recent years, the study of topological insulators (TI) in both 2D and
three dimensions (3D) has become a major focus of condensed matter physics
\cite{Qi2011,hasan2010,bernevig2006,kane2005,kane2005a,bernevig2006a,
konig2007}.
TIs maintain time-reversal (TR) symmetry, and their
band structures are characterized by the nontrivial $Z_2$-index.
As for the 3D TIs, various materials with Bloch-wave band
structures have been realized and the stable helical surface
modes have been detected \cite{fu2007,fu2007a,moore2007,qi2008,
roy2010,hsieh2008,zhang2009,hsieh2009,xia2009}.
Since LL wavefunctions have explicit forms with elegant analytical
properties, TIs based on high dimensional LL structures
would provide a nice platform for further theoretical studies.
In particular, interaction effects in the flat LLs are non-perturbative,
which could lead to non-trivial many-body states in
high dimensions.

The seminal work by Zhang and Hu \cite{zhang2001} generalizes LLs
to the compact $S^4$-sphere with particles coupled to the
$SU(2)$ gauge potential.
The isospin of particles $I$ scales  as $R^2$ where $R$ is the radius
of the sphere.
Such a system realizes the four dimensional integer and fractional TIs.
The 3D and 2D TIs can be constructed from the 4D TIs by dimensional reduction
\cite{qi2008}.
Further generalizations to other manifold have also been developed
\cite{elvang2003,bernevig2003,hasebe2010,karabali2002,nair2004,fabinger2002}.
Recently, the LLs of non-relativistic fermions have been generalized
to arbitrary dimensional flat space $R^D$ by two of the authors
\cite{li2011}.
For the simplest case of 3D,  the $SU(2)$ Aharanov-Casher gauge
potential replaces the role of the usual $U(1)$ vector potential.
Depending on the sign of the coupling constant, the flat LLs are
characterized by either positive or negative helicity.
In the positive and negative helicity channels, the eigenvalues of
spin-orbit coupling term $\vec \sigma \cdot \vec L$ take values
of $l$ and $-(l+1)$, respectively.
Each LL contributes a branch of helical surface modes at the open boundary.
When odd numbers of LLs are fully filled, there are odd numbers of
helical Fermi surfaces.
Thus the system is a 3D strong topological insulator.
This construction can be easily generalized to arbitrary $D$-dimensions
by coupling the fundamental spinors to the $SO(D)$ gauge potential.

Quantized LLs of 2D Dirac fermions have also been extensively investigated
in the field theory context known as the parity anomaly \cite{niemi1983,
jackiw1984,semenoff1984,redlich1984,redlich1984a, Fradkin1986,haldane1988}.
This can be viewed as the square root problem of the usual 2D
non-relativistic LLs.
External magnetic fields induce vacuum charges with the density proportional
to the field strength.
The sign of the charge density is related to the sign of the fermion mass.
There is an ambiguity if the Dirac fermions are massless.
In this case, there appear a branch of zero energy Landau levels.
Each of them contributes $\pm\frac{1}{2}$ fermion charge.
It is similar to the soliton charge in the Jackiw-Rebbi model
\cite{jackiw1976,niemi1986}, which is realized
in condensed matter systems of one dimensional conducting
polymer \cite{heeger1988}.
Depending on whether the zero energy Landau levels are fully
occupied or empty, the vacuum charge density is $\pm \frac{1}{2\pi l^{\prime 2}}$
where $l^{\prime}$ is the magnetic length.
In condensed matter physics, the best known example of Dirac
fermions is in graphene, which realizes a pair of Dirac cones.
The quantized LLs in graphene have been observed which distribute
symmetrically with respect to zero energy.
Their energies scale as the square root of the Landau level index.
The observed Hall conductance per spin component are quantized
at odd integer values, which reflects the nature of two
Dirac cones in graphene for each spin component
\cite{novoselov2005,zhang2005,castro_neto2009, Goerbig2011}.

In this article, we generalize the LLs with full rotational symmetry
of Dirac fermions to the three dimensional flat space and above.
It is  a square root problem of the high dimensional LLs investigated
in Ref. [\onlinecite{li2011}].
Our Hamiltonian is very simple: replacing the momentum operator in the
Dirac equation by the creation or annihilation phonon operators, which
are complex combinations of momenta and coordinates.
The LLs exhibit the same spectra as those in the 2D case but with
the full rotational symmetry in $D$-dimensional space.
Again the zero energy Landau levels are half-fermion modes.
Each LL contributes to a branch of helical surface mode at open boundaries.

This paper is organized as follows.
In Sect. \ref{sect:3DLL}, after a brief review of the 2D LL Hamiltonian of
Dirac fermions in graphene, we construct the 3D LL Hamiltonian of Dirac
fermions. Reducing this 3D system to 2D, it gives rise to 2D quantum spin
Hall Hamiltonian of Dirac fermions with LLs.
In Sect. \ref{sect:3DLLspecedge}, we further solve this 3D LL Hamiltonian of
Dirac fermions, and its edge properties are discussed.
For the later discussion of generalizing the 3D LL Hamiltonian to arbitrary
higher dimensions, in Sect. \ref{sect:so_d}, we briefly review some properties
of $D$ dimensional spherical harmonics and spinors.
In Sect. \ref{sect:oddDim} and Sect. \ref{sect:evenDim}, we extend the
solutions of LL Hamiltonians to arbitrary odd and even dimensions,
respectively.
Conclusions are given in Sect. \ref{sect:conc}.

%%%%%%%%%%%%%%%%%%%%%%%%%%%%%%%%%%%%%%%%%%%%%%%%%%%%%%%%%%%%%%%%%%%%%%%%

%*****************************************************************8
\section{The Landau level Hamiltonian of 3D Dirac fermions}
\label{sect:3DLL}

\subsection{A Brief Review of the 2D LL Hamiltonian}
Before discussing the LL problem of Dirac fermions in 3D, we briefly
review the familiar 2D case \cite{castro_neto2009, Goerbig2011} to gain
the insight on how to generalize it to high dimensions.
The celebrated condensed matter system to realize 2D Dirac fermion
is the monolayer of graphene \cite{castro_neto2009, Goerbig2011}, which possesses
a pair of Dirac cones with spin degeneracy.
Here for simplicity, we only consider a single 2D Dirac cone
under a uniform magnetic field $B \hat z$.
The Landau level Hamiltonian in the $xy$-plane reads
\bea
H_{2D,LL}=v_F \Big\{(p_x - \frac{e}{c} A_x) \sigma_x+
(p_y - \frac{e}{c} A_y) \sigma_y \Big\},
\label{eq:2DLL}
\eea
where the Dirac fermion with momentum $\vec p$ is minimally coupled to the
$U(1)$ magnetic field with symmetric gauge potentials
\bea
A_x=-\frac{B}{2}y, \ \ \, A_y=\frac{B}{2}x,
\eea
satisfying $\nabla \times \vec A = B \hat z$; the Fermi velocity $v_F$
is related to the cyclotron frequency $\omega$ via the magnetic length
$l^\prime$ as
\bea
l^\prime = \sqrt{\frac{\hbar c}{eB}}, \ \ \, v_F=\frac{l^\prime\omega}{\sqrt2}.
\eea
For later convenience, we define $l_0=\sqrt 2 l^\prime$ which will
be termed as cyclotron length below.
The spectra of Eq. \ref{eq:2DLL} consist of a branch of zero energy
LL, and other LLs with positive and negative energies distribute
symmetrically around zero energy.
The energy of each LL scales as the square root of the Landau level
index.
It is well-known that Eq. \ref{eq:2DLL} can be recast in term of creation
and annihilation operators
\bea
H_{2D,LL}=\frac{ \hbar \omega}{\sqrt 2} \left[
\begin{array}{cc}
0& \hat a_y^\dagger + i\hat a_x^\dagger \\
\hat a_y -i \hat a_x & 0
\end{array}
\right],
\label{eq:2DLL_harm}
\eea
where $\hat a_i (i=x,y)$ are  the phonon annihilation operators along the
$x$ and $y$-directions, with the form as
\bea
\hat a_i = \frac{1}{\sqrt 2}
\Big \{\frac{1}{ l_0} r_i+ i \frac{l_0}{\hbar}p_i \Big\}.
\eea

In Eq. \ref{eq:2DLL_harm}, two sets of creation and annihilation
operators combine with $1$ and the imaginary unit $i$.
In order to generalize to 3D, in which there exist three sets
of creation and annihilation operators, we employ Pauli matrices
to match them as explained below.

%-----------------------------------------------------------
\subsection{The construction of the 3D LL Hamiltonian}
We define the rotationally invariant operator $\hat B$ as
\bea
\hat B_{3D}=-i \sigma_i \hat a_i
= \sigma_i \frac{1}{\sqrt2}\Big \{ \frac{p_il_0}{\hbar }-i \frac{r_i}{l_0}
\Big\},
\label{eq:B_3D}
\eea
where the repeated index $i$ runs over $x,y$ and $z$;
$\hat a_i$ is the phonon annihilation operator along the $i$-direction;
$l_0$ is the cyclotron length.
We design the 3D Landau level Hamiltonian of Dirac fermions as
\bea
H_{3D}&=&\frac{\hbar \omega}{2}
\left[
\begin{array}{cc}
0 & B_{3D}^\dagger \\
B_{3D} & 0
\end{array}
\right ].
\label{eq:Ham_3D}
\eea
Eq. \ref{eq:Ham_3D} contains the complex combination of
momenta and coordinates, thus it can be viewed as the generalized
Dirac equation defined in the phase space.
Using the convention of $\alpha$, $\beta$ and $\gamma$-matrices defined as
\bea
\alpha_i&=&\left[
\begin{array}{cc}
0& \sigma_i \\
\sigma_i & 0 \\
\end{array}
\right], ~
\beta=\left[
\begin{array}{cc}
I_{2\times 2}& 0 \\
0& -I_{2\times 2} \\
\end{array}
\right], \nn \\
\gamma_i&=&\beta\alpha_i =\left[
\begin{array}{cc}
0& \sigma_i \\
-\sigma_i& 0 \\
\end{array}
\right],  \nn \\
\gamma_5&=&i\gamma_0\gamma_1\gamma_2\gamma_3
=\left[
\begin{array}{cc}
0& I_{2\times 2} \nn \\
I_{2\times 2} & 0
\end{array}\right],
\eea
Eq. \ref{eq:Ham_3D} is represented as
\bea
H_{3D}=v_F \sum_{i=x,y,z} \Big\{ \alpha_i  p_i
+ \gamma_i  i \hbar \frac{r_i}{l_0^2}  \Big\},
\label{eq:3D}
\eea
where $v_F= \frac{1}{2} l_0\omega$.
A mass term can be added into Eq. \ref{eq:3D} as
\bea
H_{3D,ms}= \Delta  \beta=\left( \begin{array}{cc}
\Delta I_{2\times 2}& 0 \\
0& -\Delta I_{2\times 2} \\
\end{array}
\right).
\label{eq:mass}
\eea
A similar Hamiltonian was studied before under the name of Dirac
oscillator \cite{moshinsky1989,benitez1990}, which corresponds
to Eq. \ref{eq:Ham_3D} plus the mass term of Eq. \ref{eq:mass}
with the special relation $l_0=\sqrt{\hbar c^2/\Delta \omega}$.
However, the relation between the solution of such a Hamiltonian
to the LLs and its topological properties were not noticed before.

The corresponding Lagrangian of Eq. \ref{eq:3D} reads
\bea
{\cal L}= \bar \psi \Big\{ \gamma_0 i\hbar \partial_t
- i v \gamma_i  \hbar \partial_i\Big\} \psi
- v_F \hbar  \bar\psi  i \gamma_0 \gamma_i \psi
F^{0i}(r),
\label{eq:lang}
\eea
where $F^{0i}= x_i/l_0^2$.
%Compared with the usual way of Dirac fermions coupling to the $U(1)$
%gauge field, the background field in Eq. \ref{eq:lang} couples to
%$i\gamma_0\gamma_i$.
%It can be viewed as a certain type of Pauli coupling, which
%is a non-minimal coupling of the Pauli type.
Compared with the usual way that Dirac fermions minimally couple to the $U(1)$
gauge field, here they couple to the background field in Eq. \ref{eq:lang} through
$i\gamma_0\gamma_i$.
It can be viewed as a type of non-minimal coupling, the Pauli coupling.
Apparently, Eq. \ref{eq:Ham_3D} is rotationally invariant.
It is also time-reversal invariant, and
the time-reversal operation $T$ is defined as
\bea
T=\gamma_1\gamma_3K = \left(
\begin{array}{cc}
i\sigma_2 & 0 \\
0 & i\sigma_2
\end{array}
\right) K,
\eea
where $K$ represents the complex conjugation operation,
and $T^2=-1$.

%The charge conjugation is defined as
%\bea
%C= i\gamma_2 K,
%\eea
%and the parity operation is defined as
%\bea
%P= \gamma_0
%\eea

%-------------------------------------------------------------------
\subsection{Reduction to the 2D quantum spin Hall Hamiltonian
of Dirac fermions with LLs}

If we suppress the $z$-component part in the definition of Eq. \ref{eq:B_3D},
we will arrive at double copies of the usual LL problem of 2D Dirac fermions
with Kramer degeneracy, which can be considered as the $Z_2$-topological
insulator Hamiltonian arising from LLs of 2D Dirac fermions.
We define the operator $\hat B_{2D}$ as $\hat B_{2D}= - i\sigma_x
\hat a_x -i \sigma_y \hat a_y$, and the Eq. \ref{eq:Ham_3D} reduces to
\bea
&&\frac{\sqrt 2}{2} \hbar \omega
\left(
\begin{array}{cccc}
0&0&0& \hat a_y^\dagger +i \hat a_x^\dagger\\
0&0& -\hat a_y^\dagger + i \hat a_x^\dagger &0\\
0&-\hat a_y-i \hat a_x &0&0\\
\hat a_y-i \hat a_x &0&0&0
\end{array}
\right), \nn \\
&=&v_F \left(
\begin{array}{cccc}
0&0&0& p_--A_- \\
0&0&p_++A_+ \\
0&p_-+A_-&0&0\\
p_+-A_+&0&0&0 \nn \\
\end{array}
\right),
\eea
where $p_\pm =p_x\pm  i p_y$ and $A_\pm=A_x\pm i A_y$.
It is reducible into a pair of $2\times 2$ matrices as
\bea
H_{2D,\pm}= v_F
\left(
\begin{array}{cc}
0& p_- \pm  A_-\\
p_+ \pm A_+ &0
\end{array}
\right),
\label{eq:2D}
\eea
which are time-reversal partner to each other.
Thus Eq. \ref{eq:2D} can be viewed as
the quantum spin Hall Hamiltonian of 2D Dirac fermions.

A similar situation occurs in the strained graphene systems
in which lattice distortions behave like a gauge field coupling.
Signatures of LLs due to strains have been observed in Ref. \onlinecite{Levy2010}.
Due to the TR symmetry, the Dirac cones at two non-equivalent vertices of
the Brillouin zone see gauge fields with a opposite sign to each other.
Such a coupling is also spin-independent.
However, the TR transformation connecting two Dirac cones satisfies $T^2=1$,
thus LLs due to strain are not topologically protected.
They are unstable under inter-valley scattering.

Eq. \ref{eq:2D} exhibits the standard minimal coupling to the
background $U(1)$ gauge field.
Its solutions are well-known thus will not be repeated here.
After all, there is no non-minimal coupling in 2D.
Each state of the zero energy LL is actually a half-fermion
zero mode.
Whether it is filled or empty contributes the fermion charge
$\pm\frac{1}{2}$.
As the chemical potential $\mu=0^{\pm}$, magnetic field pumps vacuum
charge density %$\rho(r)=\pm \frac{1}{2}\frac{e^2}{h}B$.
$\rho(r)=\pm \frac{1}{2}\frac{e}{hc}B$.
In the field theory context, this is an example of the parity anomaly
 \cite{niemi1983,jackiw1984,semenoff1984,redlich1984,redlich1984a,
Fradkin1986, haldane1988}.
Our 3D version and generalizations to
arbitrary dimensions exhibit similar effects as will
be discussed below.

%%%%%%%%%%%%%%%%%%%%%%%%%%%%%%%%%%%%%%%%%%%%%%%%%%%%%%%%%%%%%%%%%%%%%

\section{The bulk spectra of the 3D Dirac fermion LLs}
\label{sect:3DLLspecedge}
In this section, we will present the solution of the spectra and
wavefunctions of the 3D LL Hamiltonian for Dirac fermions, which
can be obtained basing on the solutions of the 3D LL problem
of the non-relativistic case \cite{li2011}.
We start with a brief review of the non-relativistic case of
the 3D LL problem.

\subsection{The 3D isotropic non-relativistic LL wavefunctions}
\label{subsect:review3D}

The 3D isotropic LL Hamiltonians for non-relativistic particles are just
spin-$\frac{1}{2}$ fermions in the 3D harmonic
oscillator plus spin-orbit coupling \cite{li2011} as
\bea
H_{3D,\mp}=\frac{p^2}{2M}+\frac{1}{2}M\omega^2 \mp \omega \vec L
\cdot \vec \sigma.
\label{eq:3D_nonLLL}
\eea
Their eigenfunctions are essentially the same as those of the 3D
harmonic oscillator of spin-$\frac{1}{2}$ fermions organized in the
total angular momentum eigen-basis of $j, j_z$ as
\bea
\psi_{n_r,j_\pm, l,j_z}(\vec r)=R_{n_r,l}(r){\cal Y}_{j_\pm,l,j_z}
(\hat\Omega),
\label{eq:harmo}
\eea
where $n_r$ is the radial quantum number;
$j_\pm=l\pm \frac{1}{2}$ represent positive and negative helicity channels,
respectively; and $l$ is the orbital angular momentum.
Please note that $l$ is not an independent variable from $j_\pm$.
We write it explicitly in order to keep track of the orbital angular
momentum.
The radial wavefunction can be represented through the
confluent hypergeometric functions as
\bea
R_{n_r,l}(r)=N_{n_r,l} \Big(\frac{r}{l_0}\Big)^l
F(-n_r,l+\frac{3}{2},\frac{r^2}{l_0^2}) e^{-\frac{r^2}{2l_0^2}},
\eea
where $F$ is the standard first kind confluent hypergeometric function,
\bea
F(-n_r,l+\frac{3}{2},\frac{r^2}{l_0^2})&=&\sum_{n=0}^{\infty}
\frac{\Gamma(-n_r +n)}{\Gamma(-n_r)}
\frac{\Gamma(l+\frac{3}{2})} {\Gamma(l+\frac{3}{2}+n)}
\nn \\
&\times& \frac{1}{\Gamma(n+1)}\left(\frac{r^2}{l_0^2}\right)^n.
\eea
When $n_r$ is a positive integer, the sum over $n$ is cut off
at $n_r$.
The normalization factor reads as
\bea
N_{n_r,l}=\frac{l_0^{-\frac{3}{2}}}{ \Gamma(l+\frac{3}{2})}
\sqrt{\frac{2\Gamma(l+n_r+\frac{3}{2})}{\Gamma(n_r+1)}}.
\eea

The angular part of the wavefunction is the standard spin-orbit coupled
spinor spherical harmonic function, which reads as
\bea
{\cal Y}_{j_\pm,l,j_z=m+\frac{1}{2}}(\hat \Omega) =\left(
\begin{array}{c}
\pm \sqrt{\frac{l\pm j_z+\frac{1}{2}}{2l+1}} Y_{l,m} (\hat \Omega)\\
\sqrt{\frac{l\mp j_z+\frac{1}{2}}{2l+1}} Y_{l,m+1} (\hat \Omega)
\end{array}
\right).
\eea

Depending on the sign of the spin-orbit coupling in Eq. \ref{eq:3D_nonLLL},
one of the two branches of positive or negative  helicity states
are dispersionless with respect to $j$, and thus are dispersionless LLs.
For  $H_{3D,-}$, the positive helicity states become dispersionless LLs
as
\bea
H_{3D,-} \psi_{n_r,j_+, l, j_z}= (2n_r+\frac{3}{2}) \hbar\omega
\psi_{n_r,j_+, l, j_z}
\eea
where $n_r$ serves as Landau level index.
However, the negative helicity states are dispersive whose eigen-equation
reads
\bea
H_{3D,-} \psi_{n_r,j_-, l, j_z}= (2n_r+ 2l + \frac{5}{2})\hbar \omega
\psi_{n_r,j_-, l, j_z}.
\eea
Similarly, we have the following eigen-equations for the $H_{3D,+}$ as
\bea
H_{3D,+} \psi_{n_r,j_-, l, j_z} &=& (2n_r+\frac{1}{2})\hbar\omega
\psi_{n_r,j_-, l, j_z} \nn \\
H_{3D,+} \psi_{n_r,j_+, l, j_z}&=& (2n_r+2l +\frac{3}{2})\hbar\omega
\psi_{n_r,j_+, l, j_z}.
\eea
In this case, the negative helicity states become dispersionless
LLs with respect to $j$, while the positive ones are dispersive.

\subsection{3D LL wavefunctions of Dirac fermions}

Now we are ready to present the spectra and the four-component eigenfunctions
of Eq. \ref{eq:3D} for the massless case.
Its square is block-diagonal, and two blocks become the
non-relativistic 3D Landau level Hamiltonians with opposite
signs of spin-orbit coupling studied in Ref. [\onlinecite{li2011}],
\bea
\frac{H^2_{3D}}{\frac{1}{2}\hbar \omega}&=&
\left[\begin{array}{cc}
H_-& 0 \\
0& H_+\\
\end{array}
\right ] \nn \\
&=&
\frac{p^2 }{2M}  + \frac{1}{2}M \omega^2 r^2
-\omega \Big \{\vec L \cdot \vec \sigma +\frac{3}{2}\hbar \Big \}
\left[ \begin{array}{cc}
I& 0 \\
0 & -I \\
\end{array}
\right],  \nn \\
\label{eq:Hsquare}
\eea
where $M$ is defined through the relation
$l_0=\sqrt{\hbar/(M\omega)}$.

Its eigenfunctions can be represented in terms of
non-relativistic Landau levels of Eq. \ref{eq:harmo}
as presented in Sect. \ref{subsect:review3D}.
Eq. \ref{eq:Ham_3D} has a conserved quantity as
\bea
K=\left[
\begin{array}{cc}
\vec \sigma \cdot \vec l +\hbar& 0 \\
0 & -(\vec \sigma\cdot \vec l +\hbar)
\end{array}
\right].
\eea
According to its eigenvalues, the eigenfunctions of Eq. \ref{eq:Ham_3D}
are classified
as
\bea
K\Psi^I_{\pm n_r, j, j_z}&=&(l+1)\hbar \Psi^I_{\pm n_r, j, j_z},\nn \\
K\Psi^{II}_{\pm n_r, j, j_z}&=&-l\hbar \Psi^{II}_{\pm n_r, j, j_z},
\eea
respectively.
$\Psi^{I}_{\pm n_r, j, j_z}$ is dispersionless with respect to $j$,
while $\Psi^{II}_{\pm n_r, j, j_z}$ is dispersive, respectively.
The dispersionless branch is solved as
\bea
\Psi_{\pm n_r,j, j_z}^I (\vec r)=\frac{1}{\sqrt 2}
\left [
\begin{array}{c}
    \psi_{n_r,j_+, l, j_z} (\vec r)\\
\pm i\psi_{n_r-1,j_-, l+1, j_z }(\vec r)
\end{array}
\right]
\label{eq:LLWF}
\eea
with the energy
\bea
E_{\pm n_r, j, j_z}=\pm \hbar \omega \sqrt n_r.
\eea
Please note that the upper and lower two components  of Eq. \ref{eq:LLWF}
possess different values of orbital angular momenta.
They exhibit opposite helicities of $j_\pm$, respectively.
The zeroth Landau level ($n_r=0$) is special: only the first two components
are non-zero.

On the other hand, the wavefunctions of the dispersive branch read
\bea
\Psi_{\pm n_r,j, j_z}^{II} (\vec r)=\frac{1}{\sqrt 2}
\left [
\begin{array}{c}
\mp i    \psi_{n_r,j_-,l+1, j_z} (\vec r)\\
    \psi_{n_r ,j_+, l, j_z} (\vec r)
\end{array}
\right],
\label{eq:dispersive}
\eea
with the spectra solved as
\bea
E_{\pm n_r, j, j_z}=\pm \hbar  \omega \sqrt{n_r +j +1}.
\eea
These states are just discrete energy levels lying between two adjacent
LLs.
For simplicity, let us only consider the positive energy states.
The degeneracy of these mid-gap states lying between the $n$-th and
$(n+1)$-th Landau levels with $n=n_r+j+\frac{1}{2}$ is finite, $n(n+1)$, due
to finite combinations of $n_r$ and $j$.
In particular, between the zeroth LL and the LLs with $n_r=\pm 1$, these
discrete states do not exist at all.

Because Eq. \ref{eq:3D} satisfies $\beta H_{3D} \beta =-H_{3D}$,
its spectra are symmetric with respect to zero energy.
If the zeroth branch of Landau levels $(n_r=0)$ are occupied,
each of them contribute a half-fermion charge.
The vacuum charge is $\rho^{3D}(\vec r)=\frac{1}{2} \sum_{j,j_z}
\Psi^\dagger_{0;j,j_z} (\vec r) \Psi_{0;j,j_z}(\vec r)$, which are
calculated as
\bea
\rho^{3D} (\vec r) &=& \frac{1}{2l_0^3} \Big\{ \sum_{l=0}^\infty \frac{l+1}{2\pi}
\frac{1}{\Gamma(l+\frac{3}{2})}\Big(\frac{r^2}{l_0^2}\Big)^l
\Big\}
e^{-\frac{r^2}{l_0^2}}
\nn \\
&=& \frac{1}{2\pi l_0^3}
\Big\{ \frac{1}{\sqrt \pi} e^{-\frac{r^2}{l_0^2}} +\big (\frac{r}{l_0}
+\frac{l_0}{2r}\big) \mbox{erf}\big(\frac{r}{l_0}\big) \Big\}, \nn \\
&\longrightarrow& \frac{r}{2\pi l_0^4}, \mbox{~~as~~} r\rightarrow +\infty.
\label{eq:3D_density}
\eea

In 2D, the induced vacuum charge density in the gapless Dirac LL
problem is a constant, $\rho^{2D}(r)=\frac{1}{2 \pi l_0^2} =
\frac{1}{4\pi} \frac{e}{\hbar c} B$ , which is known as ``parity anomaly''.
However, in the 3D case, the vacuum charge density $\rho^{3D}(r)$
diverges linearly, which is dramatically different from that in 2D.
This can be easily understood in the semi-classic picture. Each
Landau level with orbital angular momentum $l$ has a classic radius
$r_l=\sqrt{2 l} l_0$. In 2D, between $r_l<r<r_{l+1}$, there is only
one state. However, in 3D there is the $2j_++1=2l+2$ fold degeneracy,
which is the origin of the divergence of the vacuum charge density as $r$
approaches infinity. Generally speaking, in the case of $D$ dimensions,
the degeneracy density scales as $r^{D-2}$ as shown in Sect.
\ref{sect:oddDim} and Sect.\ref{sect:evenDim}.
The intrinsic difference between high-$D$ and 2D is that
the high dimensional LL problems exhibit the form of non-minimal
coupling. In 2D, due to the specialty of Pauli matrices, this kind of
coupling reduces back to the usual minimal coupling.
In Eq. \ref{eq:lang}, the background field is actually a linear divergent
electric field, not the magnetic vector potential.
Eq. \ref{eq:3D_density} can be viewed as a generalization of
``parity anomaly'' to 3D for non-minimal couplings.

Now we consider the full Hamiltonian with the mass term Eq. \ref{eq:mass}.
The mass term mixes the LLs in Eq. \ref{eq:LLWF} with opposite
level indices $\pm n_r$ but the same values of $j$ and $j_z$.
The new eigenfunctions become
\bea
\left[
\begin{array}{c}
\Psi_{n_r,j, j_z}^{I,\prime}\\
\Psi_{-n_r,j,j_z}^{I,\prime}
\end{array}
\right]=
\left[
\begin{array}{cc}
\cos\theta & -\sin\theta \\
\sin\theta & \cos\theta
\end{array}
\right ]
\left[
\begin{array}{c}
\Psi_{n_r,j, j_z}\\
\Psi_{-n_r,j,j_z}
\end{array}
\right],
\eea
where $\cos^2\theta =\frac{1}{2}\Big [1+ \sqrt n_r/\sqrt{ n_r +[\Delta/
(\hbar \omega)]^2} \Big]$.
The spectra are
\bea
E^{I,ms}_{\pm n_r,j,j_z}=\pm \sqrt {n_r (\hbar \omega)^2
+\Delta^2}.
\eea
The zeroth LL $\Psi^I_{n_r=0,j,j_z} (\vec r)$ singles out,
which is not affected by the mass term.
Only its energy is shifted to $\Delta$.

%--------------------------------------------------------------------
\subsection{Gapless surface modes}
\label{sect:surface}

As shown in Eq. \ref{eq:3D_density}, the 3D LL system of Dirac fermions
has a center, and does not have translational symmetry.
Thus how to calculate its topological index remains a challenging
problem.
Nevertheless, we can still demonstrate its non-trivial topological
properties through the solution of its gapless surface modes.

We consider the surface spectra at a spherical boundary with a
large radius $R\gg l_0$.
The Hamiltonian $H_{r<R}$ inside the sphere takes the
massless form of Eq. \ref{eq:3D},
while $H_{r>R}$ outside takes the mass term of Eq. \ref{eq:mass}
in the limit of $|\Delta|\rightarrow \infty $.
Again the square of this Hamiltonian $(H_{r<R}+H_{r>R})^2$ is
just Eq. \ref{eq:Hsquare} subject to the open boundary condition
at the radius of $R$.
%The squared Hamiltonian Eq. \ref{eq:Hsquare} has the supersymmetric
%structure \cite{bagchi2001}, in which $H_\mp$ form a supersymmetric partner.
%The lowest LL is singly degeneracy with respect the radial
%quantum number $n_r=0$, and other LLs are doubly degenerate.
%Such a structure remains exact for the open boundary problem
%since the boundary condition is homogeneous.
The spectra of the open surface problem of the non-relativistic
3D LL Hamiltonian have been calculated and presented in Fig. 3
in Ref. [\onlinecite{li2011}], in which the spectra of
each Landau level remain flat for bulk states and
develop upturn dispersions as increasing $j$ near the surface.
The solution to the Dirac spectra is just to take the square root.
Except the zeroth LL, each of the non-relativistic
LL and its surface branch split into a pair of bulk and surface branches
in the relativistic case.
The relativistic spectra take the positive and negative square roots of the
non-relativistic spectra, respectively.
The zeroth LL branch singles out.
We can only take either the positive or negative square root, but not both.
It surface spectra are upturn or downturn with respect to
$j$ depending on the sign of the vacuum mass.
For the current Hamiltonian, only the first two components of
the zeroth LL wavefunction are non-zero, thus it only senses
the upper $2\times 2$ diagonal block of vacuum mass in Eq. \ref{eq:mass},
thus its surface spectra are pushed upturn.
%The relation between these two open boundary problems for the
%non-relativistic and relativistic LL problems is sketched in
%Fig. \ref{fig:surface}.

%-------------------------------------------------------------------
%\begin{figure}[tbp]
%\centering
%\epsfig{file=surface.eps,clip=2,width=\linewidth, height=0.5
%\linewidth,angle=0}
%\caption{The sketched spectra of the open surface problem with the radius
%$R\gg l_0$ for (a) the non-relativistic LL Hamiltonian Eq. \ref{eq:Hsquare}
%and (b) the relativistic LL Hamiltonian.
%Each pair of LL branches in (a) for $n_r\ge 1$ splits into a pair
%of both positive and negative square-root branches in (b), whose
%surface spectra become upturn and downturn, respectively.
%For the LLL branch in (a), either positive or negative square root
%is taken for (b) depending on the sign of the vacuum mass.
%}
%\label{fig:surface}
%\end{figure}
%------------------------------------------------------------------

%%%%%%%%%%%%%%%%%%%%%%%%%%%%%%%%%%%%%%%%%%%%%%%%%%%%%%%%%%%%%%%%%%%
\section{Review of D-dimensional spherical harmonics and spinors}
\label{sect:so_d}
We will study the LL of Dirac fermions for general dimensions in
the rest part of the paper.
For later convenience, we present here some background knowledge of
the $SO(D)$ group which can found in standard group theory textbooks
\cite{hamermesh1989}.

The $D$-dimensional spherical harmonic functions
$Y_{l,\{m\}}(\hat\Omega)$ form the representation of the $SO(D)$ group
with the one-row Young pattern, where $l$ is the number of boxes and
$\{ m \}$ represents a set of $D-2$ quantum numbers of the
subgroup chain from $SO(D-1)$ down to $SO(2)$.
The degeneracy of $Y_{l;\{m\}}$ is
\bea
d_{[l]}(SO(D))=(D+2l-2)  \frac{(D+l-3)!}{l!(D-2)!}.
\eea
Its Casimir is $\sum_{i<j} L^2_{ij}=l(l+D-2)\hbar^2$, where
the orbital angular momenta are defined as $L_{ij}=r_i p_j -r_j p_i$.

We also need to employ the $\Gamma$-matrices.
The $2\times 2$ Pauli matrices are just the rank-1 $\Gamma$-matrices.
They can be generalized to rank-$k$ $\Gamma$-matrices which
contains $2k+1$ matrices anti-commuting with each other.
Their dimensions are  $2^k\times 2^k$.
A convenient recursive definition is constructed based on
the rank-$(k-1)$ $\Gamma$-matrices as
\bea
\Gamma^{(k)}_i&=&\left[
\begin{array}{cc}
0& \Gamma_a^{(k-1)}\\
\Gamma_a^{(k-1)}& 0
\end{array}
\right], \ \ \
\Gamma^{(k)}_{2k}=\left[
\begin{array}{cc}
0& -i I \\
iI & 0
\end{array}
\right], \nn \\
\Gamma^{(k)}_{2k+1}&=&
\left[
\begin{array}{cc}
I& 0 \\
0& -I
\end{array}
\right],
\eea
where $i=1,..., 2k-1$.
For $D=2k+1$-dimensional space, its fundamental spinor is $2^k$-dimensional.
The generators are constructed
$S_{ij}=\frac{1}{2}\Gamma_{ij}^{(k)}$
where
\bea
\Gamma^{(k)}_{ij}=-\frac{i}{2}[\Gamma^{(k)}_i, \Gamma_j^{(k)}].
\eea
For the $D=2k$-dimensional space, there are two irreducible fundamental
representations with $2^{k-1}$ components.
Their generators are denoted as $S_{ij}$ and $S_{ij}^\prime$, respectively,
which can be constructed based on both rank-$(k-1)$
$\Gamma_i^{(k-1)}$ and $\Gamma_{ij}^{(k-1)}$-matrices.
For the first $2k-1$ dimensions, the generators share the same form as
\bea
S_{ij}=S^\prime_{ij}=\frac{1}{2}\Gamma_{ij}^{(k-1)}, ~~ 1\le i < j \le 2k-1,
\label{eq:so2k_1}
\eea
while other generators $S_{i,2k}$ and $S_{i,2k}^\prime$
differ by a sign as
\bea
S_{i,2k}=S^\prime_{i,2k}=
\pm \frac{1}{2}\Gamma_{i}^{(k-1)}, ~~1\le i \le 2k-1.
\label{eq:so2k_2}
\eea

We couple $Y_{l,\{m\}}$ to the $SO(D)$ fundamental irreducible spinors.
For simplicity, we use the same symbol $s$ in this paragraph
to denote the fundamental spinor representation (Rep.) for $SO(D)$ with $D=2k+1$ and
the two irreducible spinor Reps. for $SO(D)$ with $D=2k$.
The states split into the positive $(j_+)$ and
negative $(j_-)$ helicity sectors.
The bases are expressed as ${\cal Y}_{j_\pm;s,l,\{j_m\}}(\hat \Omega)$,
where $\{j_m\}$ is a set of $D-2$ quantum numbers for the
subgroup chain.
The degeneracy number of ${\cal Y}_{j_+;s,l,\{j_m\}}(\hat \Omega)$ is
\bea
d_{j_+}=d_s \frac{(D+l-2)!}{l! (D-2)!},
\label{eq:dim_j+}
\eea
where $d_s$ is the dimension of the fundamental spinor representation.
Similarly, the degeneracy number of ${\cal Y}_{j_-,;s,l,\{j_m\}}(\hat \Omega)$ is
\bea
d_{j_-}=d_s \frac{(D+l-3)!}{(l-1)! (D-2)!}.
\eea

The eigenvalues of the spin-orbit coupling term $\sum_{i<j}\Gamma_{ij} L_{ij}$
for the sectors of ${\cal Y}_{j_+;s,l,\{j_m\}}(\hat \Omega)$
and ${\cal Y}_{j_-;s,l,\{j_m\}}(\hat \Omega)$ are $l\hbar$ and $-(l+D-2)\hbar$,
respectively.
We present the eigenstates of the $D$-dimensional
harmonic oscillator with fundamental spinors in the total angular
momentum basis as
\bea
\psi^D_{n_r,j_\pm, s, l, \{j_m\} }(\vec r)=R_{n_r,l}(r){\cal Y}_{j_\pm,s,l,\{j_m\}}
(\hat\Omega),
\label{eq:ND_harmo}
\eea
where the radial wavefunction reads
\bea
R_{n_{r},l}(r)= N^D_{n_r,l} \left( \frac{r}{l_0}\right)
^{l}e^{-\frac{r^{2}}{2l_{g}^{2}}}F(-n_{r},l+\frac{D}{2},\frac{r^{2}}
{l_{0}^{2}})
\eea
and  the normalization constant reads
\bea
N^D_{n_r,l}=\frac{l_0^{-\frac{D}{2}}}{ \Gamma(l+\frac{1}{2}D)}
\sqrt{\frac{2\Gamma (n_{r}+l+\frac{D}{2})}{\Gamma(n_r+1)}}.
\eea

%%%%%%%%%%%%%%%%%%%%%%%%%%%%%%%%%%%%%%%%%%%%%%%%%%%%%%%%%%%%%%%%%%%%

\section{The LLs of odd dimensional Dirac fermions }
\label{sect:oddDim}
In this section, we generalize the 3D LL Hamiltonian for Dirac fermions
to an arbitrary odd spatial dimensions $D=2k+1$.
We need to use the rank-$k$ $\Gamma$-matrices, which contains
$2k+1$ anti-commutable matrices at the dimensions of $2^k\times 2^k$
denoted as $\Gamma_i^{(k)} (1\le i\le 2k+1)$.
The definition of $\Gamma_i^{(k)}$ and the background information
of the representation of the $SO(D)$ group is given in
Sect. \ref{sect:so_d}.

We define $\hat{B}_{2k+1}=-i\Gamma_i^{(k)} \hat{a}_i$ and
the $2k+1$-dimensional LL Hamiltonian of Dirac fermions
$H_{2k+1}$ in the same way as in Eq. \ref{eq:Ham_3D}.
Again the square of $H_{2k+1}$ reduces to a block-diagonal form as
\bea
\frac{(H_{2k+1})^2}{\frac{1}{2}\hbar \omega}&=&
\frac{p^2 }{2M}  + \frac{1}{2}M \omega^2 r^2
-\omega \Big \{\sum_{i<j} L_{ij} \Gamma_{ij}^{(k)} \nn \\
&+&\frac{2k+1}{2}\hbar \Big \}
\left[ \begin{array}{cc}
I& 0 \\
0 & -I \\
\end{array}
\right],
\label{eq:2k+1D}
\eea
where $\Gamma_{ij}^{(k)}=-\frac{i}{2}[\Gamma_i^{(k)},\Gamma_j^{(k)}]$.
Each diagonal block of Eq. \ref{eq:2k+1D} is just the form the
$2k+1$ D LL problem of non-relativistic fermions in Ref. [\onlinecite{li2011}].

Again we can define the following conserved quantity
\bea
K=\left[
\begin{array}{cc}
\Gamma_{ij}^{(k)} L_{ij}+(D-2)\hbar &0\\
0& -(\Gamma_{ij} L_{ij}+\hbar)
\end{array}
\right],
\label{eq:K}
\eea
$K$ divides the eigenstates into two sectors $\Psi^I_{\pm n_r, j,\{j_m\}}$
and $\Psi^{II}_{\pm n_r, j,\{ j_m \}}$,
\bea
K\Psi^I_{\pm n_r, j, \{j_m\}}&=&\hbar (l+D-2)\Psi^I_{\pm n_r, j, \{j_m\}},\nn \\
K\Psi^{II}_{\pm n_r, j, \{j_m\}}&=&- \hbar l \Psi^{II}_{\pm n_r, j, \{j_m\}},
\eea
respectively.
As explained in Sect. \ref{sect:so_d}, $j$ represents the
spin-orbit coupled representation for the $SO(D=2k+1)$ group,
and $\{j_m\}$ represents a set of good quantum number
of the subgroup chain from $SO(D-1)$ down to $SO(2)$.

Similarly as before, the sectors of $\Psi^{I,II}_{\pm n_r, j, \{j_m\}}$
are dispersionless and dispersive with respect to $j$, respectively.
The concrete wavefunctions are the same as those in
Eq. \ref{eq:LLWF} and Eq. \ref{eq:dispersive}
by replacing the 3D wavefunction to the $D$-dimensional version
of Eq. \ref{eq:ND_harmo}.
The wavefunctions of $\Psi^{I,II}_{\pm n_r, j, \{j_m\}}$ are
given explicitly as
\bea
\Psi_{\pm n_r,j, j_z}^I (\vec r)&=&\frac{1}{\sqrt 2}
\left [
\begin{array}{c}
    \psi^{D}_{n_r, s, j_+,l , \{j_m\} } (\vec r)\\
\pm i\psi^{D}_{n_r-1, s,  j_-, l+1, \{ j_m \}  }(\vec r)
%    \psi^{D}_{n_r, s, \bf{j_+=l+\frac{1}{2}}, \{j_m\} } (\vec r)\\
%\pm i\psi^{D}_{n_r-1, s,  \bf{ j_-=(l+1)-\frac{1}{2} }, \{ j_m \}  }(\vec r)
\end{array}
\right], \nn \\
\Psi_{\pm n_r,j, j_z}^{II} (\vec r)&=&\frac{1}{\sqrt 2}
\left [
\begin{array}{c}
\mp i    \psi^{D}_{n_r,s, j_-,l+1, \{ j_m \} } (\vec r)\\
    \psi^{D}_{n_r, s, j_+, l, \{j_m\}} (\vec r)
%\mp i    \psi^{D}_{n_r,s, \bf{ j_-=(l+1)-\frac{1}{2} }, \{ j_m \} } (\vec r)\\
%    \psi^{D}_{n_r, s, \bf{j_+=l+\frac{1}{2}}, \{j_m\}} (\vec r)
\end{array}
\right].
\eea
The dispersion relation for the LL branch of $\Psi^{I}_{\pm n_r, j, \{j_m\}}$
still behaves as $E_{\pm n_r; j, \{j_z\} }=\pm \hbar \omega \sqrt n_r$,
while that for the branch of $\Psi^{II}_{\pm n_r, j, \{j_m\}}$
reads as $E_{\pm n_r; j, \{j_z\} }=\pm \hbar
\omega \sqrt {n_r+l+\frac{D}{2}}$.

Again each occupied zero energy LL contributes to $\frac{1}{2}$-fermion
vacuum charge.
If the zeroth LLs are fully filled, the vacuum charge is still
$\rho^D(\vec r)=\frac{1}{2}\sum_{j,\{j_m\}} |\Psi^I_{0,j,\{j_m\}} (\vec r)|^2$, which
is expressed as
\bea
\rho^D(r)=\frac{1}{l_0^{D}} \Big\{\sum_{l=0}^{\infty}
\frac{1}{\Gamma(l+\frac{D}{2})} \Big(\frac{r}{l_0}\Big)^{2l}
\frac{g_l (D)}{\Omega_D}
 \Big\}
e^{-\frac{r^2}{l_0^2}},
\label{eq:D_density}
\eea
where $D=2k+1$; $g_l(D)$ is the degeneracy of the positive helicity
sector of the fundamental spinor coupling to the $l$-th $D$-dimensional
spherical harmonics, and its expression is the same
as $d_{j_+}$ given in Eq. \ref{eq:dim_j+};
$\Omega_D=D\pi^{D/2}/\Gamma(D/2+1)$ is the area of $D$-dimensional
unit sphere.
Eq. \ref{eq:D_density} can be summed analytically as
\bea
\rho(r)&=&\frac{\sqrt 2}{4}\Big(\frac{2}{\pi l_0^2}\Big)^{\frac{D}{2}}
F(D-1, \frac{D}{2},\frac{r^2}{l_0^2}) e^{-\frac{r}{l_0^2}}\nn \\
&\longrightarrow& \frac{1}{(2\pi)^{\frac{D-1}{2}} l_0^D}
\frac{1}{\Gamma(\frac{D-1}{2})} \Big( \frac{r}{l_0 }\Big)^{D-2},
\mbox{~as~~} r\rightarrow \infty. \nn \\
\eea

Similarly, if the $D$-dimensional version of the mass term inside
Eq. \ref{eq:mass} is added, every wavefunction with the radial
quantum number $n_r$ hybridizes with its partner with $-n_r$
while keeping all other quantum numbers the same.
The pair of new eigenvalues becomes $\pm \sqrt{(\hbar \omega)^2 n_r+\Delta^2}$.
Again, the zero-th LL wavefunctions single out and remain the same,
but their energies are shifted to $\Delta$.
For a similar open surface problem to that in Sect. \ref{sect:surface},
each LL with $n_r\neq 0$ develops a branch of gapless surface
mode with the upturn (downturn) dispersion with respect to $j$ for $n_r>0$
($n_r<0$), respectively.
The surface mode from the zeroth LL develops either upturn
or downturn dispersions depending on the relative sign
of the background field coupling and the vacuum mass.

%%%%%%%%%%%%%%%%%%%%%%%%%%%%%%%%%%%%%%%%%%%%%%%%%%%%%%%%%%%%%%
\section{ The LLs of even dimensional Dirac fermions}
\label{sect:evenDim}
The LL problem in the even dimensions with $D=2k$ is more
complicated.
The $SO(2k)$ group has two irreducible fundamental spinor
representations $s$ and $s^\prime$.
Each of them is with the dimension of $2^{k-1}$.
The construction of the $SO(2k)$ generators for the irreducible
representations are introduced in Sect. \ref{sect:so_d}.

Now we define $\hat B_{2k}=-i\Gamma^{(k)}_i \hat a_i$ where $i$
runs over $1$ to $2k$.
Similarly to the 2D case, the counterpart of Eq. \ref{eq:Ham_3D} in
the $D=2k$ dimensions $H_{D=2k}$ is reducible into two $2^k\times 2^k$
blocks as
\bea
H_\pm =\frac{\hbar \omega}{2}\left[ \begin{array}{cc}
0& \pm \hat a^\dagger_{2k} + i \Gamma_i^{(k-1)} \hat a_i^\dagger\\
\pm \hat a_{2k} - i \Gamma_i^{(k-1)} \hat a_i &0
\end{array}
\right], \ \ \,
\label{eq:reduce}
\eea
where the repeated index $i$ runs over from $1$ to $2k-1$.
For each one of the reduced Hamiltonians $H_\pm$,
each off-diagonal block has only the $SO(2k-1)$ symmetry.
Nevertheless, each of $H_\pm$ is still $SO(2k)$ invariant.
If we combine the two irreducible fundamental spinor representations
$s$ and $s^\prime$ together, the spin generators are defined as
\bea
S_{ij;s\oplus s^\prime}=-\frac{i}{4}[\Gamma_i^{(k)}, \Gamma_j^{(k)}].
\eea
Both of $H_{\pm}$ commute with the total angular momentum
operators in the combined representation of $s\oplus s^\prime$
defined as
\bea
J_{ij;s\oplus s^\prime}=L_{ij}+S_{ij;s\oplus s^\prime}.
\eea

We choose $H_+$ as an example to present the solutions of the LL
wavefunctions in even dimensions.
The $K$-operator is similarly defined as in Eq. \ref{eq:K} as
\bea
K_+=\left[
\begin{array}{cc}
2S_{ij} L_{ij}+(D-2)\hbar &0\\
0& -(2S^\prime_{ij} L_{ij}+\hbar)
\end{array}
\right],
\label{eq:K_even}
\eea
where $i,j$ run from 1 to $2k$, and $S_{ij}$ and $S^\prime_{ij}$
are generators in the two fundamental spinor representations
given in Eqs. \ref{eq:so2k_1} and \ref{eq:so2k_2}, respectively.
They again can be divided into two sectors of $\Psi^{+,I}$ and $\Psi^{+,II}$
whose eigenvalues of $K_+$ are $\hbar (l+D-2)$ and $-\hbar l$, respectively.
The dispersionless branch of $\Psi^{+,I}$ can be viewed as LLs, whose
wavefunctions read
\bea
\Psi_{\pm n_r, j, \{j_m\}}^{+,I} (\vec r)=\frac{1}{\sqrt 2}
\left [
\begin{array}{c}
         \psi_{n_r,j_+,s, l, \{j_m\}} (\vec r)\\
\mp i    \psi_{n_r-1 ,j_-,  s^\prime, l+1, \{j_m\}} (\vec r)
\end{array}
\right].
\label{eq:LLevenD_1}
\eea
Their spectra are same as before $E_{\pm n_r, j, \{j_m\}}=\pm  \hbar
\omega \sqrt {n_r}$.
Please note that the upper and lower components involve the $s$ and
$s^\prime$ representations, respectively.
Similarly, the dispersive solutions of $\Psi^{II}$ become
\bea
\Psi_{\pm n_r, j, \{j_m\}}^{+,II} (\vec r)=\frac{1}{\sqrt 2}
\left [
\begin{array}{c}
\mp i   \psi_{n_r, j_-, s, l+1, \{j_m\}} (\vec r)\\
        \psi_{n_r, j_+,  s^\prime, l, \{j_m\}} (\vec r)
\end{array}
\right],
\label{eq:LLevenD_2}
\eea
whose dispersions read $E_{\pm n_r, j, \{j_m\}}=\pm  \hbar
\omega \sqrt {n_r+l+\frac{D}{2}}$.
The solutions to $H_-$ are very similar to Eq. \ref{eq:LLevenD_1}
and Eq. \ref{eq:LLevenD_2} by exchanging the irreducible
fundamental spinor representation indices $s$ and $s^\prime$.

Again if the zeroth branch LLs are filled, the vacuum charge
$\rho^{D}(\vec r)=\frac{1}{2}\sum_{j,\{j_m\}} |\Psi^I_{0,j,\{j_m\}} (\vec r)|^2$
is calculated as
\bea
\rho^{D}(r)&=&\frac{1}{l_0^{D}} \Big\{\sum_{l=0}^{\infty}
\frac{1}{\Gamma(l+\frac{D}{2})} \Big(\frac{r}{l_0}\Big)^{2l}
\frac{g_l (D)}{\Omega_D}
 \Big\}
e^{-\frac{r^2}{l_0^2}}, \ \ \,
\eea
where $D=2k$, $\Omega_D$ and $g_l(D)$ are defined similarly
as in Eq. \ref{eq:D_density}.
It can be summed over analytically as
\bea
\rho^{D}(r)
&=&\frac{1}{4} \Big (\frac{2}{\pi l_0}\Big)^{\frac{D}{2}}
F(D-1,\frac{D}{2}, \frac{r^2}{l_0^2}) e^{-\frac{r^2}{l_0^2}},
\eea
which $\rho^D(r)
\longrightarrow
 \frac{\sqrt \pi }{(2\pi l_0^2)^{\frac{D}{2}}}
\frac{1}{\Gamma(\frac{D-1}{2})} \Big( \frac{r}{l_0 }\Big)^{D-2}$,
as $r\rightarrow +\infty$.

%%%%%%%%%%%%%%%%%%%%%%%%%%%%%%%%%%%%%%%%%%%%%%%%%%%%%%%%%%%%%%%%
\section{Conclusion}
\label{sect:conc}

In summary, we have generalized the LL problem of 2D Dirac fermions
to arbitrary higher dimensional flat spaces with spherical symmetry.
This problem is essentially the square root problem of its non-relativistic
LL problem with spherical symmetry in high dimensions Ref. \onlinecite{li2011}.
The zero energy LLs is a branch of $\frac{1}{2}$-fermion modes.
On the open boundary, each LL contributes one branch of helical surface
modes.
This series of LL problems can be viewed as the generalization of parity
anomaly in 2D to arbitrary dimensions in a spherical way.
An open question is that how to experimentally realize the case
of the 3D systems.

{\it Note added} \ \ \
Near the completion of this manuscript, we became
aware of that Eq. \ref{eq:3D} plus Eq. \ref{eq:mass} with a special
relation between the background field coupling constant and the mass term
was studied in Ref. [\onlinecite{moshinsky1989, benitez1990}]
under the name of Dirac oscillator.
We have studied a more general form from a different perspective
and identified their relation to the LLs.

This work was supported in part by the NBRPC (973 program) 2011CBA00300
(2011CBA00302) and Grant No. ARO-W911NF0810291.
K. I. was supported by DOE-FG03-97ER40546.

%%%%%%%%%%%%%%%%%%%%%%%%%%%%%%%%%%%%%%%%%%%%%%%%%%%%%%%%%%%%%%%%%%%
%%%%%%%%%%%%%%%%%%%%%%%%%%%%%%%%%%%%%%%%%%%%%%%%%%%%%%%%%%%%%%%%%%%
%%%%%%%%%%%%%%%%%%%%%%%%%%%%%%%%%%%%%%%%%%%%%%%%%%%%%%%%%%%%%%%%%%%

%\appendix

%\bibliographystyle{prsty}
%\bibliography{TI,TI_exp,topo}

\begin{thebibliography}{10}

\bibitem{li2011}
Y. {Li} and C. {Wu}, ArXiv:1103.5422  (2011).

\bibitem{thouless1982}
D.~J. Thouless, M. Kohmoto, M.~P. Nightingale, and M. den~Nijs,
Phys. Rev. Lett. {\bf 49},  405  (1982).

\bibitem{kohmoto1985}
M. Kohmoto,
Ann. Phys. {\bf 160},  343  (1985).

\bibitem{haldane1988}
F.~D.~M. Haldane,
Phys. Rev. Lett. {\bf 61},  2015  (1988).

\bibitem{Qi2011}
X.~L. Qi and S.~C. Zhang,
Rev. Mod. Phys. {\bf 83}, 1057 (2011).

\bibitem{hasan2010}
M. Hasan and C. Kane,
Rev. Mod. Phys. {\bf 82},  3045  (2010).

\bibitem{bernevig2006}
B. Bernevig, T. Hughes, and S. Zhang,
Science {\bf 314},  1757  (2006).

\bibitem{kane2005}
C.~L. Kane and E.~J. Mele,
Phys. Rev. Lett. {\bf 95},  146802  (2005).

\bibitem{kane2005a}
C.~L. Kane and E.~J. Mele,
Phys. Rev. Lett. {\bf 95},  226801  (2005).

\bibitem{bernevig2006a}
B.~A. Bernevig and S.~C. Zhang,
Phys. Rev. Lett. {\bf 96},  106802  (2006).

\bibitem{konig2007}
M. K\"onig, S. Wiedmann, C. Br\"une, A. Roth, H. Buhmann, L. Molenkamp, X.~L. Qi, S.~C. Zhang,
Science {\bf 318},  766  (2007).

\bibitem{fu2007}
L. Fu and C.~L. Kane,
Phys. Rev. B {\bf 76},  045302  (2007).

\bibitem{fu2007a}
L. Fu, C.~L. Kane, and E.~J. Mele,
Phys. Rev. Lett. {\bf 98},  106803  (2007).

\bibitem{moore2007}
J.~E. Moore and L. Balents,
Phys. Rev. B {\bf 75},  121306  (2007).

\bibitem{qi2008}
X.~L. Qi, T.~L. Hughes, and S.~C. Zhang,
Phys. Rev. B {\bf 78},  195424  (2008).

\bibitem{roy2010}
R. Roy,
New J. Phys. {\bf 12},  065009  (2010).

\bibitem{hsieh2008}
D. Hsieh, D. Qian, L. Wray, Y. Xia, Y. Hor, R. Cava, and M. Hasan,
Nature {\bf 452},  970  (2008).

\bibitem{zhang2009}
H. Zhang, C. Liu, X.~L. Qi, X. Dai, Z. Fang, and S.~C. Zhang,
Nature Phys. {\bf 5},  438  (2009).

\bibitem{hsieh2009}
D. Hsieh, Y. Xia, D. Qian, L. Wray, F. Meier, J.~H. Dil, J. Osterwalder, L. Patthey, A.~V. Fedorov, H. Lin, A. Bansil, D. Grauer, Y.~S. Hor, R.~J. Cava, and M.~Z. Hasan,
Phys. Rev. Lett. {\bf 103},  146401  (2009).

\bibitem{xia2009}
Y. Xia, D. Qian, D. Hsieh, L. Wray, A. Pal, H. Lin, A. Bansil, D. Grauer, Y. Hor, R. Cava, and M. Z. Hasan, Nature Phys. {\bf 5},  398  (2009).

\bibitem{zhang2001}
S. Zhang and J. Hu,
Science {\bf 294},  823  (2001).

\bibitem{elvang2003}
H. Elvang and J. Polchinski,
Comptes Rendus Physique {\bf 4},  405  (2003).

\bibitem{bernevig2003}
B.~A. Bernevig, J. Hu, N. Toumbas, and S.~C. Zhang,
Phys. Rev. Lett. {\bf 91}, 236803  (2003).

\bibitem{hasebe2010}
K. Hasebe,
SIGMA (Symmetry, integrability and geometry: methods and applications) {\bf  6}, 071   (2010).

\bibitem{karabali2002}
D. Karabali and V. P. Nair,
Nucl. Phys. {\bf B 641}, 533 (2002).

\bibitem{nair2004}
V. P. Nair and S. Randjbar-Daemi,
Nucl. Phys. {\bf B 679}, 447 (2004).

\bibitem{fabinger2002}
M. Fabinger,
JHEP {\bf 2002}, 037  (2002).

\bibitem{niemi1983}
A.~J. Niemi and G.~W. Semenoff,
Phys. Rev. Lett. {\bf 51},  2077  (1983).

\bibitem{jackiw1984}
R. Jackiw,
Phys. Rev. D {\bf 29},  2375  (1984).

\bibitem{redlich1984}
A.~N. Redlich,
Phys. Rev. Lett. {\bf 52},  18  (1984).

\bibitem{redlich1984a}
A.~N. Redlich,
Phys. Rev. D {\bf 29},  2366  (1984).

\bibitem{semenoff1984}
G.~W. Semenoff,
Phys. Rev. Lett. {\bf 53},  2449  (1984).

\bibitem{Fradkin1986}
E. Fradkin,E. Dagotto, and D. Boyanovsky,
Phys. Rev. Lett. {\bf 57},  2967  (1986).

\bibitem{jackiw1976}
R. Jackiw and C. Rebbi,
Phys. Rev. D {\bf 13},  3398  (1976).

\bibitem{niemi1986}
A.~J. Niemi and G.~W. Semenoff,
Phys. Rep. {\bf 135},  99  (1986).


\bibitem{heeger1988}
A.~J. Heeger, S. Kivelson, J.~R. Schrieffer, and W.~P. Su,
Rev. Mod. Phys. {\bf  60},  781  (1988).

\bibitem{novoselov2005}
K.~S. Novoselov, A. Geim, S. Morozov, D. Jiang, M. Grigorieva, S. Dubonos, and  A. Firsov,
\nat {\bf 438},  197  (2005).

\bibitem{zhang2005}
Y. {Zhang}, Y.-W. {Tan}, H.~L. {Stormer}, and P. {Kim},
\nat {\bf 438},  201  (2005).

\bibitem{castro_neto2009}
A.~H. {Castro Neto}, F. Guinea, N. M. R. Peres, K. S. Novoselov, and A. K., Geim,
Rev. Mod. Phys. {\bf 81},  109  (2009).

\bibitem{Goerbig2011}
M.~O. Goerbig,
Rev. Mod. Phys. {\bf 83}, 1193 (2011).

\bibitem{moshinsky1989}
M. Moshinsky and A. Szczepaniak,
J. Phys. A {\bf 22},  L817  (1989).

\bibitem{benitez1990}
J. Benitez, R.~P. Martinez y Romero, H.~N. N\'unez-Y\'epez, and A.~L. Salas-Brito,
Phys. Rev. Lett. {\bf 64},  1643  (1990).

\bibitem{Levy2010}
N. Levy, S.~A. Burke, K.~L. Meaker, M. Panlasigui, A. Zettl, F. Guinea, A.~H. Castro Neto, and M.~F. Crommie,
Science {\bf 329}, 544 (2010).

%\bibitem{bagchi2001}
%B. Bagchi, {\em {Supersymmetry in quantum and classical mechanics}} %(Chapman \&
%  Hall/CRC, ADDRESS, 2001).

\bibitem{hamermesh1989}
M. Hamermesh, {\em Group Theory and Its Application to Physical Problems}
  (Dover, New York, 1989).

\end{thebibliography}

\end{document}